# NbodyCP: A direct N-body simulation code for composite stellar populations of single and binary star clusters

Zhong-Mu Li[1] and Rainer Spurzem[2;3;4]

[1]  Institute of Astronomy, Dali University, Dali 671003, PR China; *zhongmuli@126.com*

[2]  Zentrum fu"r Astronomie, Astronomisches Rechen-Institut, Mo"nchhofstr. 12–14, Heidelberg 69120, Germany

[3]  National Astronomical Observatories, Chinese Academy of Sciences, 20A Datun Rd., Chaoyang District, 100101, Beijing, China

[4]  Kavli Institute for Astronomy and Astrophysics, Peking University, 5 Yi He Yuan Road, Haidian District, Beijing 100871, P.R. China

**Abstract** It is well-known that some star clusters contain composite stellar populations (CSPs), in which the metallicities or (and) ages of stars are different. The formation and evolution of such clusters and their stellar populations remain unclear. Both single and binary cluster channels may lead to such CSPs. In order to simulate the formation and evolution of such CSPs in star clusters, this work develops a code of direct N-body simulation of CSPs, NbodyCP. It is applied to different clusters, in particular, to binary clusters. It shows that CSPs and different kinds of cluster pairs can be formed via dynamical processes. This will help to partially explain the formation of CSPs and various clusters. Some special cluster structures, e.g., two cores or bar-like shape, are shown to be the results of evolution of some binary clusters. The simulation also shows that the separation between the members of a binary cluster affects the time of two member clusters to combine or move away significantly.

## 1  INTRODUCTION

Composite stellar populations (CSPs) or multiple stellar populations (MSPs) are widely observed in globular clusters. Many such clusters exhibit extended main sequence turnoffs (eMSTOs) and multiple sequences in the color-magnitude diagrams (CMDs) (e.g., Goudfrooij et al. 2011a,b; Milone et al. 2015; D'Antona et al. 2015; Brandt & Huang 2015; Li et al. 2016, 2017a). A similar phenomenon has also been observed in a few open clusters (Cordoni et al. 2018; Marino et al. 2018; Bastian et al. 2018; Gossage et al. 2019;

Sun et al. 2019; Piatti & Bonatto 2019; de Juan Ovelar et al. 2020; Lipatov et al. 2022). Although some different factors are thought to be the possible reasons of the eMSTOs of star clusters, there is no certain answer yet. Usually, the age spread, chemical abundance differences and variations of stellar rotation rates are thought to be the most probable factors. Age spread ($300 \sim 500$ Myr) plays a key role in the study of CSPs of star clusters because it can explain most eMSTO phenomena (Li et al. 2016), and more and more chemical abundance studies showed that there are two or more generations of stars in many clusters (Marino et al. 2009, 2019; Milone et al. 2020). If star clusters had already formed at the very beginning and evolved independently as individual clusters, stellar rotation seems the main reason for eMSTOs.. However, if it has undergone mergers during the process of cluster evolution, the eMSTOs may result from the evolution of binary clusters.

In fact, many star cluster pairs have been found. Some of them show different ages for their member clusters. For example, Kovaleva et al. (2020) found a pair of open clusters, Collinder 135 and UBC 7. Camargo (2021) reported a pair of open clusters that may have different ages (600 Myr and 2 Gyr for NGC1605a and NGC1605b respectively). We do not know the formation of such clusters and their stellar populations clearly. N-body simulation will help to understand the formation and evolution of such cluster pairs and their stellar populations better. There have been some previous works. Arnold et al. (2017) did a very simple simulation about the formation of cluster pairs within the first 20 Myr, but they did not take into account some important factors such as stellar evolution and external tidal fields. Recently, Darma et al. (2021) did a similar work and suggested that binary star clusters can form from stellar aggregates with a variety of initial conditions, but binary-star evolution was not taken into account. N-body simulations of single star clusters with MSP are not very common, but see the notable exception cited below.

In this paper we use the state-of-the-art direct force integration N-body integration code for star clusters NBODY6++GPU, which is optimised for parallelisation using simultaneously three different levels: the bottom-end is many core GPU accelerated parallel computing (Nitadori & Aarseth 2012; Wang et al. 2015), mid-level OpenMP thread based parallelisation, and upper level MPI parallelisation (Spurzem 1999). It yields excellent sustained performance on current hybrid massively parallel supercomputers (Spurzem & Kamlah 2023). It is a successor to the many direct force integration N-body codes of gravitational N-body problems, which were originally written by Sverre Aarseth (Aarseth 1985, 1999a,b, 2003; Aarseth et al. 2008, and sources therein). The recent review by Spurzem & Kamlah (2023) provides a comprehensive overview over the entire research field of collisional dynamics and how NBODY6++GPU fits within this field. That review also provides some key informations about and citations of recent competitor codes, such as PeTaR (Wang et al. 2020) and BiFROST (Rantala et al. 2023). In Kamlah et al. (2022) many more details can be found about the substantial improvement of SSE and BSE (single and binary stellar evolution) as compared to their initial state (Hurley et al. 2005, and earlier references therein) - also in that paper the level of agreement with SSE and BSE used in Monte Carlo simulations (Hypki et al. 2025; Hypki & Giersz 2013) is discussed. Monte Carlo codes have been used intensively with large numbers of models to look for the dynamic and kinematic evolution of two distinct populations (MSP). Starting with initially more concentrated second population (and different maximum mass) the mixing of populations for single and binary stars up to the present day were simulated and compared with observations (Giersz et al. 2025; Livernois et al. 2024; Hypki et al. 2022; Hong et al. 2017). Monte Carlo simulations, while being computationally very fast, use some strong approximations, such as strict spherical symmetry of the gravitational potential, and the relaxation and encounter cross sections modelled in a statistical way. There is only one remarkable sequence of papers published forr N-body simulations with two stellar populations (Hong et al. 2015, 2016, 2017, 2019). While these N-body models could be comparable to the MOCCA models, they are modelling only very small single star clusters (like $N \approx 20.000$), not merging clusters.

This paper presents a direct N-body simulation code for both single and binary star clusters, NbodyCP, in which CSPs and SSPs can be included. The NbodyCP code is developed from the Nbody6++GPU[1]. This ensures the reliability of code. It is then used to simulate the evolution of various star clusters, to study how the stellar population and cluster structure evolve.

This paper is organized as follows: the main features of NbodyCP are described in section 2. An example application of NbodyCP to the formation of CSP from high mass ratio binary cluster in section 3. An application to a low mass ratio binary cluster is given in section 4. Section 5 studies the effect of initial separation on the evolution of binary cluster. Finally, section 6 concludes and discusses on this paper.

## 2   NBODYCP CODE

In order to simulate the CSPs of star clusters, we develop a new version of the direct N-body simulation code NBODY6++GPU (Wang et al. 2015; Kamlah et al. 2022; Spurzem & Kamlah 2023). Because the NBODY6++GPU code can only simulate star cluster with SSP, in which all stars have the same metallicity and age, it can not be used directly for this work. We therefore develop the code to simulate CSPs of single and binary clusters, and call the new code NbodyCP (CP means composite population). Comparing to Nbody6++GPU, the main changes and features of NbodyCP code are as follows.

- First, the generation of initial star sample was changed. The main feature is that two groups of stars can be generated. The metallicities, ages and binary fractions of two groups can be different or the same. The centers of two groups can be set separately, so both single and binary clusters can be simulated. The upper limit of stellar mass is extended to $150M_\odot$, because some black hole binaries have been found when studying gravitational waves (Abbott et al. 2020).

- Second, the calculation mode of stellar evolution is changed, so it can evolve stars with different metallicities respectively.

- Third, the input and output formats are changed, in order to make it easy to be used. The format of input parameters are rearranged, so the inputs are easier to be set compared to the original NBODY6++GPU version. The outputs are rearranged to let one can read the text format results conveniently, but the HDF5 format results are not changed. This makes up for the defect of the NBODY6++GPU that is not friendly enough to read the results.

- Fourth, an assistance program was invented to complete the computation automatically. The reason is that the code sometimes stops for some known or unknown reasons, so it needs to reset some parameters and restart the program. Doing it manually will waste a lot of time, as we can not check the working status from time to time.

---

[1] https://github.com/nbody6ppgpu/

As a whole, the NbodyCP code is an updated version of the NBODY6++GPU, and it includes all features of the NBODY6++GPU code. In particular, some high-speed calculation technologies are used in the simulation code. Different CPU cores are called by the code via the OpenMP technology, and different servers can be called to work together via the MPI technology. GPU and processor instruction set are also used to speed up the calculation. This makes it possible to do direct N-body simulations on a small or medium computing server.

## 3  SIMULATION OF FORMATION OF CSPS FROM HIGH MASS RATIO BINARY CLUSTERS

The NbodyCP code is used to simulate the formation of CSPs of star clusters, via the merger of high mass ratio binary clusters. Here we assume that the stars of a CSP distribute originally in two component clusters and they distribute in a single cluster finally in spatial coordinates, which is similar to the observations of some GCs and OCs. Because CSPs were found in globular clusters earlier than in open clusters, we take two globular clusters in this simulation. The number of stars of two component clusters are set to the same, 250000. The mass ratio is therefore close to one. All stars are assumed to be single stars following the initial mass function ofKroupa(2001). The stars of each component cluster are distributed by the Plummer model (Plummer 1911). The metallicity of stars are set to $Z = 0.008$. The two component clusters are thought to be distant 90 pc from each other. The ages of two component clusters are set to be different by 300 Myrs.

Fig. 1 shows the spatial distributions of stars at different ages, clearly demonstrating that the distribution of stars changes a lot with the evolution of binary cluster. In particular, some strange spatial structures, e.g., bar-like core, form naturally at some ages, e.g., $107.41 - 213.38$ Myr. It indicates that some extended structures, which are observed in some clusters (Rser et al. 2019), may be related to the dynamical interactions between clusters.

Fig. 2 shows the evolution of Hertzsprung-Russell diagram (HRD) of the same binary cluster. We observe that eMSTOs structure form at some ages (e.g., a few hundred million years). This means that the eMSTOs of some clusters may result from dynamical interactions between (or among) clusters.

If the distance between two component clusters are not too large, a single cluster will form within about 3 Gyr, which is similar to the observational result of star clusters with eMSTO.

## 4  SIMULATION OF TWO CLUSTERS WITH DIFFERENT METALLICITIES AND AGES

A main feature of NbodyCP is the ability to simulate the formation and evolution of the pair of star clusters. Here we give an example of the application of NbodyCP to a pair of star clusters, whose component clusters have different star numbers (7500 and 2500), metallicities (0.008 and 0.01 for $Z$) and ages (300 and 0 Myr for the large and small components). The initial separation of two clusters is set to 50 pc.

Fig. 3 illustrates the spatial evolution of stars in the cluster pair, which comprises a massive cluster (blue points) and a less massive counterpart (red points). Although two clusters separate 50 pc from each other at the beginning, they get closer and closer. The member stars of both clusters become increasingly dispersed, and gradually blend together. After about 100 Myr, it is completely impossible to distinguish stars from different clusters based on their spatial distribution.

Fig. 4 shows the evolution of velocities of the star in Fig. 3. It is clear that the velocities of most stars become closer and closer. However, there are a very small number of stars whose velocities are significantly different from those of the vast majority of stars, at some ages. This may partially explain the observed high-speed stars Brown et al. (2005).

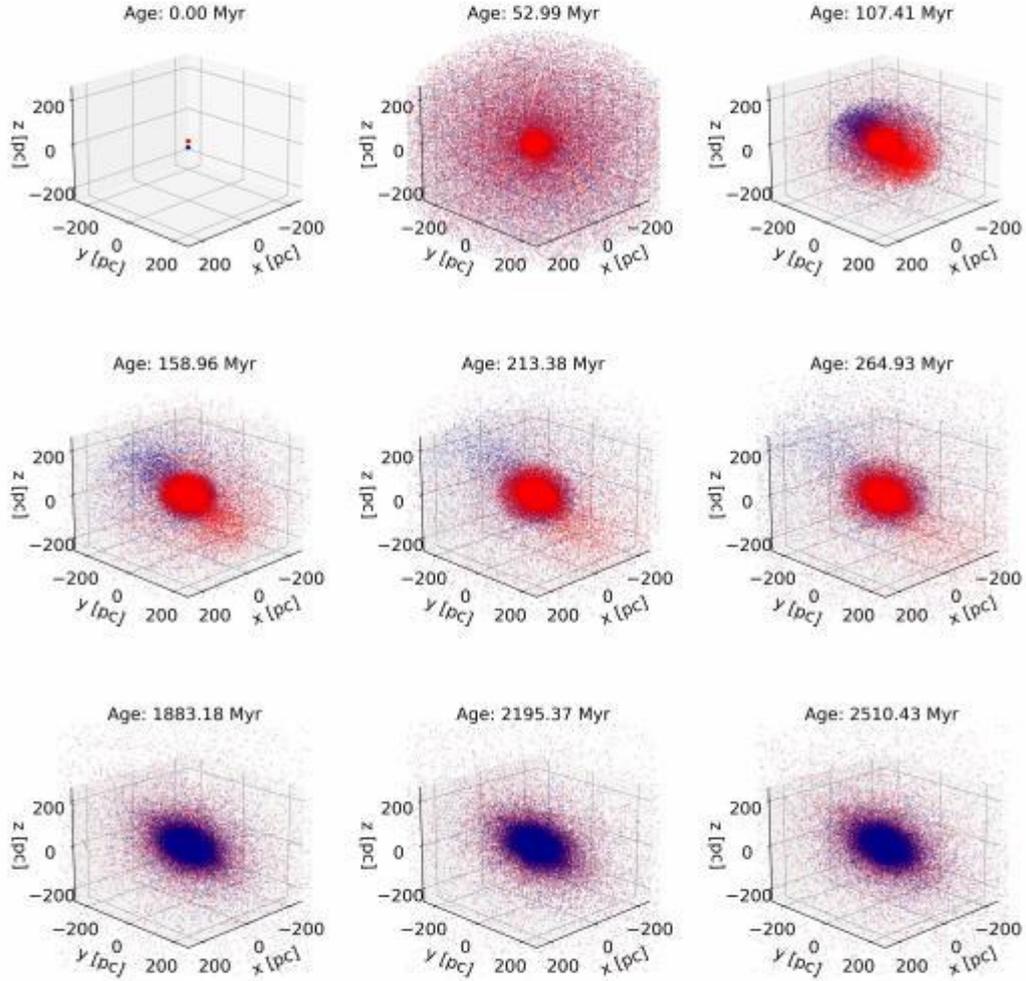

Fig. 1: Evolution of spatial distribution of a binary cluster including composite stellar populations. Two component clusters have the same star number, and they are 90 pc distant at the beginning.

# 5 EFFECT OF SEPARATION OF BINARY CLUSTERS

As another application of NbodyCP, this section studies the effect of separation of binary clusters. Here we put two star clusters at three different distances to check how their evolution can be different. The initial separations are taken as 10, 30, 50 pc. The evolution of spatial distribution of stars are given by Figs. 5, 6, and 7, respectively for three binary clusters with various separations. The results show that the two member clusters of all the three binary clusters combined into one finally, and the initial separation of two member clusters determines the time of combination process. The durations are approximately 10, 20, and 200 Myr for the combination processes of binary clusters with initial separations from 10 to 50 pc. In particular, some interesting shapes of spatial distributions of stars form in the evolution of binary clusters. For example, a cluster with two cores forms from the binary cluster with initial separation of 50 pc, at an age of about 73 Myr. In addition, a shape like a bar form at an age near 92 Myr. This implies that some simple models (e.g., Plummer model) is not necessarily enough to describe the distribution of stars when star clusters form from different sources. This is important for the deep studies of formation of star clusters.

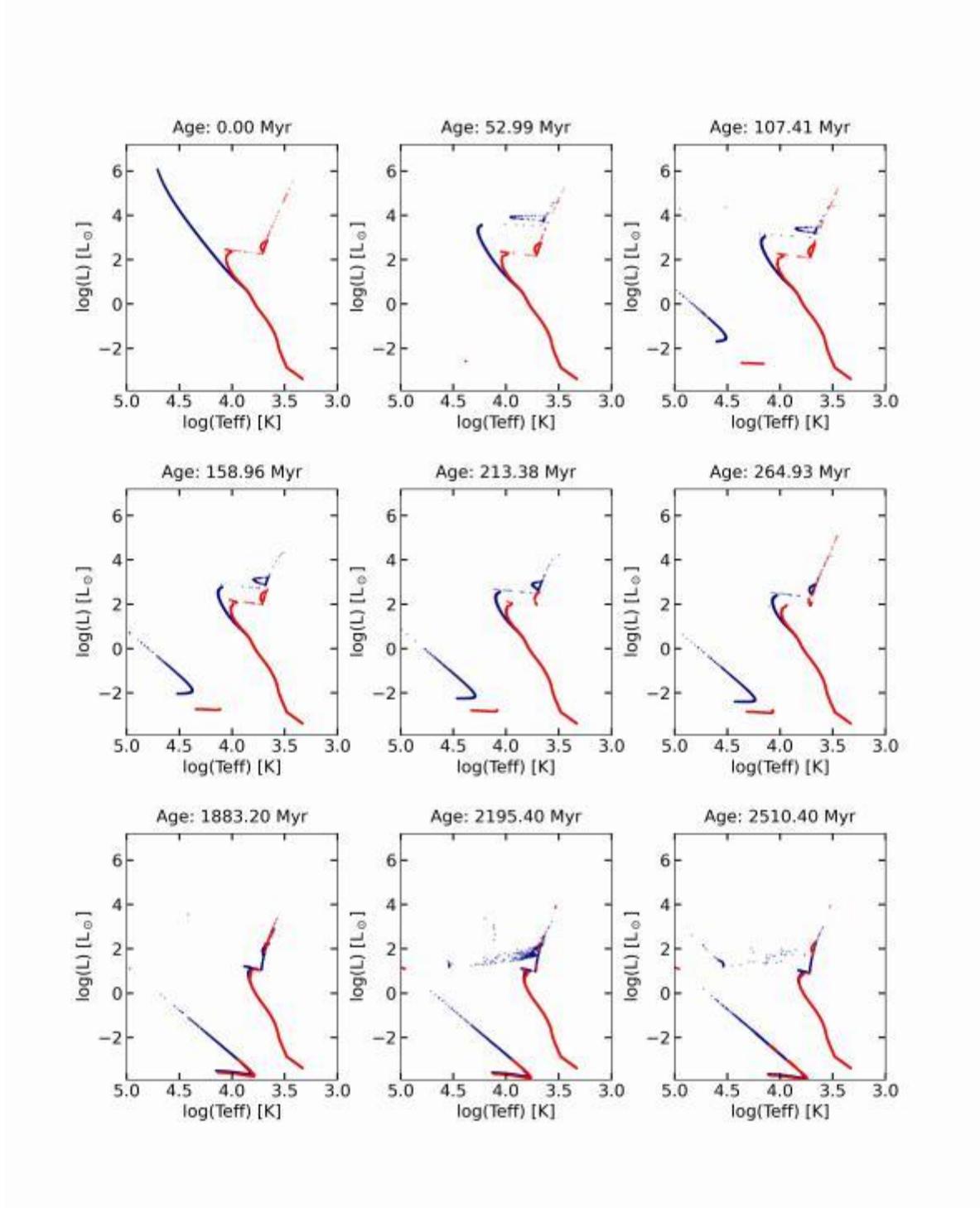

Fig. 2: Evolution of Hertzsprung-Russell diagrams of the same binary cluster in Fig. 1.

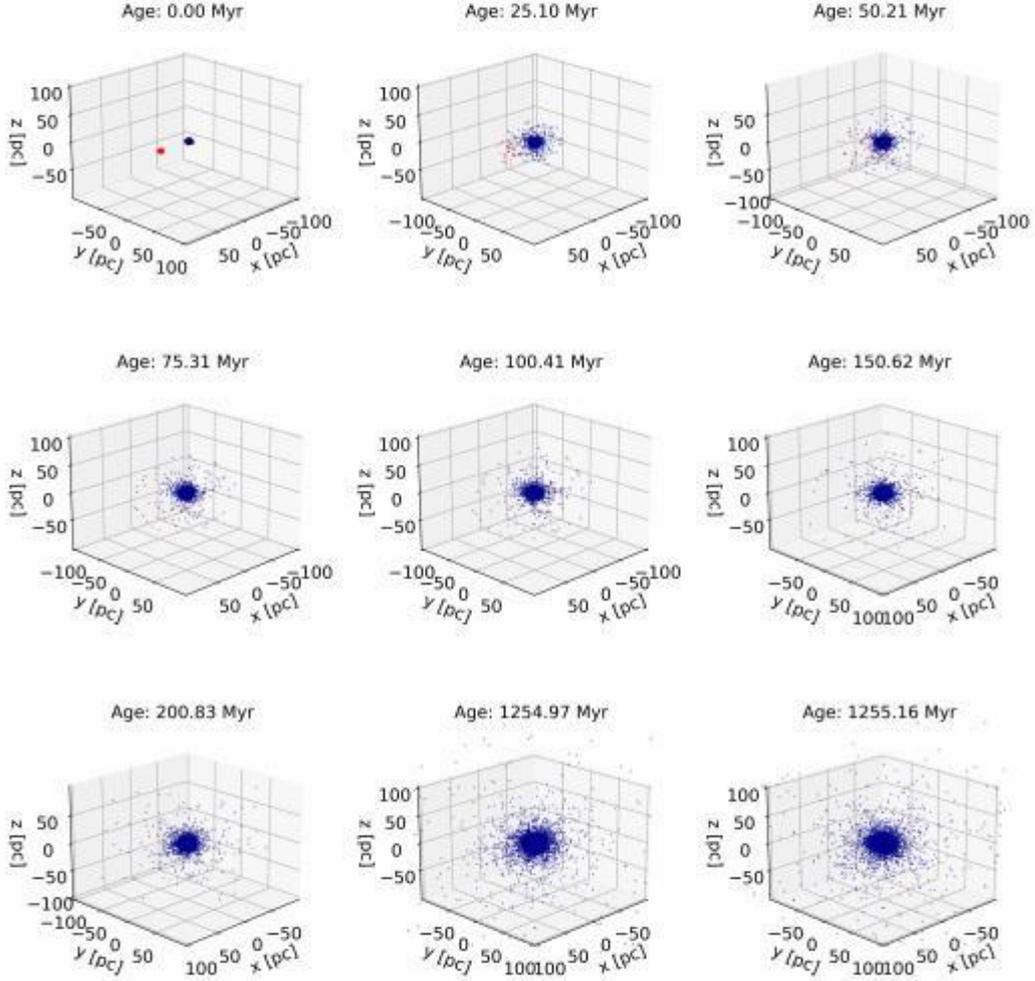

Fig. 3: Evolution of spatial distribution of stars of a low mass ratio binary cluster. Two component clusters have different metallicities and ages.

## 6  CONCLUSION AND DISCUSSION

This paper introduces a direct N-body simulation code, NbodyCP, for single and binary clusters. Both composite and simple stellar populations (CSPs and SSPs) can be included when using this code. It is useful for studying the formation and evolution of both normal and abnormal star clusters. It is also helpful for studying the formation of super massive black holes (Gond n 2023). In particular, it can be used for in-depth researches on stellar populations of star clusters. Because this code is developed based on the renowned Nbody6++GPU code, its reliability and computational accuracy is well ensured. Meanwhile, it adopts technologies such as GPU acceleration and instruction set acceleration, which makes it have a relatively fast computing speed, similar to Nbody6++GPU code.

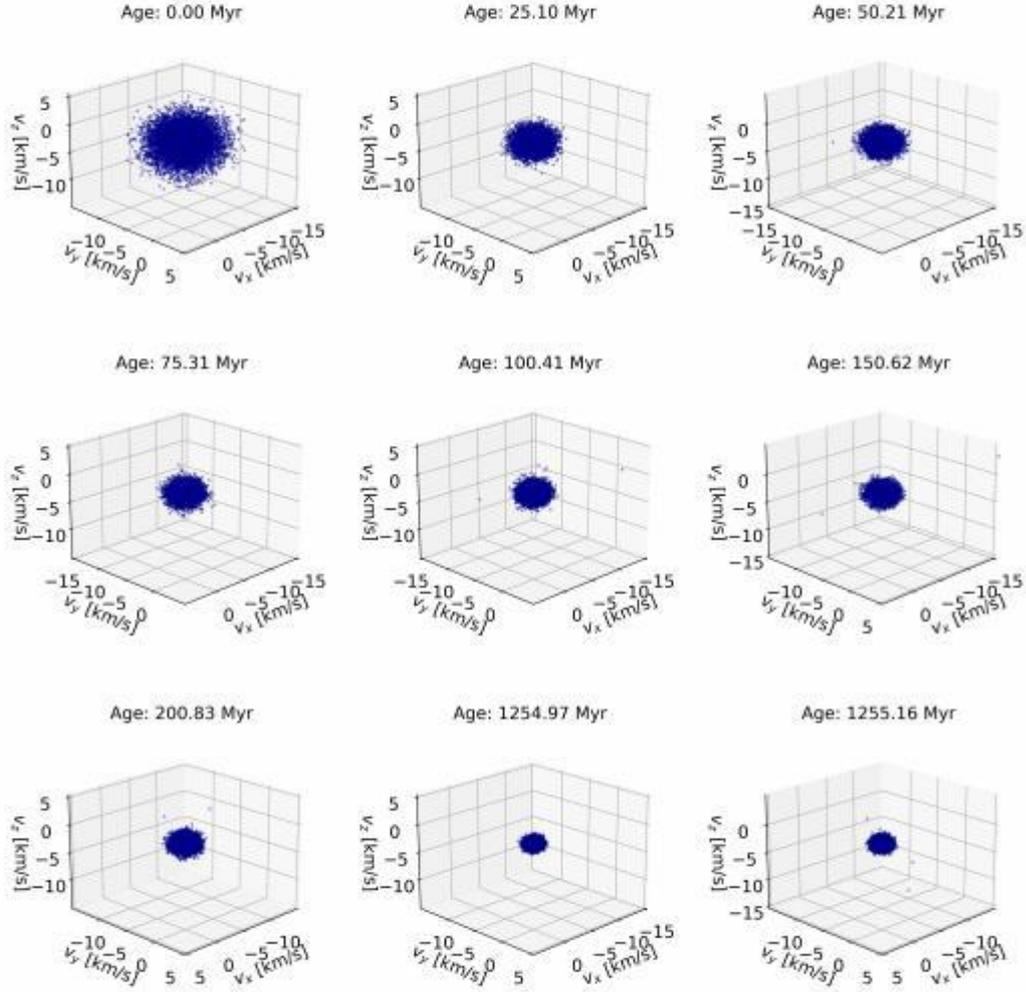

Fig. 4: Similar to Fig. 3, but for the evolution of velocities of member stars.

A few examples are given in this paper, to show how this code can be used in star cluster simulations. Some important results, e.g., the evolution of spatial distribution of stars and HRDs of the cluster are obtained. It helps us to better understand the formation of some peculiar structures (e.g., bar-like structure and cluster with two cores) of star clusters, and the formation of CSPs (or eMSTOs) of some clusters. When this code is used to compare the evolution of three binary clusters with various separations, it is shown that the time of combination of two member clusters is affected significantly by the separation between two member clusters. This suggests that NbodyCP code is of powerful ability in different kinds of studies of both single and binary star clusters.

Although this code takes many advanced computational technologies, it will take a very long period to finish the simulation if the star number of a cluster is larger than about 500 000. This means that it is better to run this code on some computing servers or clusters equipped with a large number of high-performance

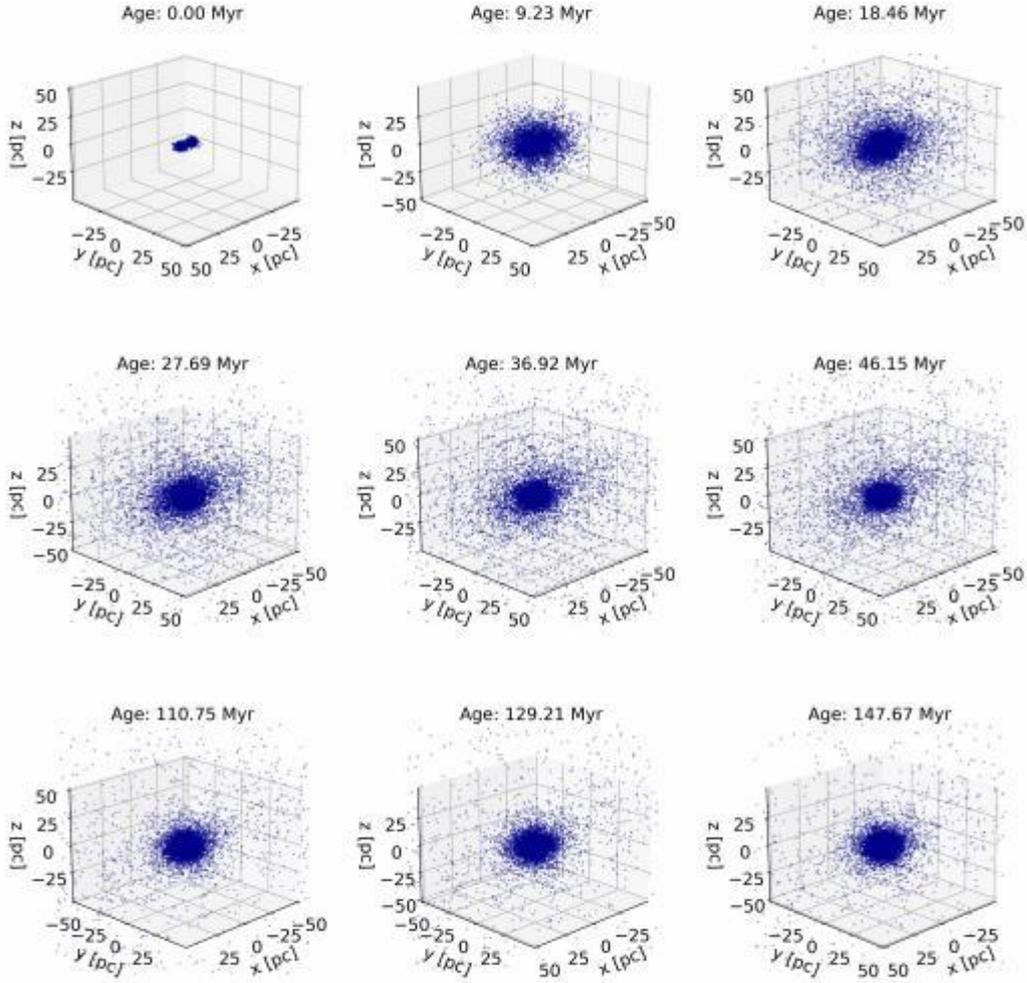

Fig. 5: Evolution of spatial distribution of stars of a close binary cluster. The initial separation is 10 pc, and star number is 10000.

CPU cores and GPUs. When the number of stars is larger than about one million, the calculation quantity is extremely large. It is difficult to do such a simulation on an ordinary PC. Even though many different technologies, e.g., MPI and GPU, are used to speed up the calculation, such simulations may require computational times longer than one month (Kamlah et al. 2022; Spurzem & Kamlah 2023). We show as an example an unpublished benchmark using Nbody6++GPU on the raven GPU cluster of Max-Planck data and computing facility in Garching, Germany (MPCDF).

Fig. 8 shows short test simulations using up to 16 million stars, using up to 16 nodes with 64 GPUs of type A100, each having about 5000 cores. To our knowledge this is the largest ever published particle number using Nbody6++GPU; but this illustrates nevertheless shows that even on such a supercomputer simulations of several Gyrs will take months. But one can also see that the speed-up due to parallelization is near ideal (diagonal) for the largest models. Dashed lines represent predictions of a timing model, which will be published in future work.

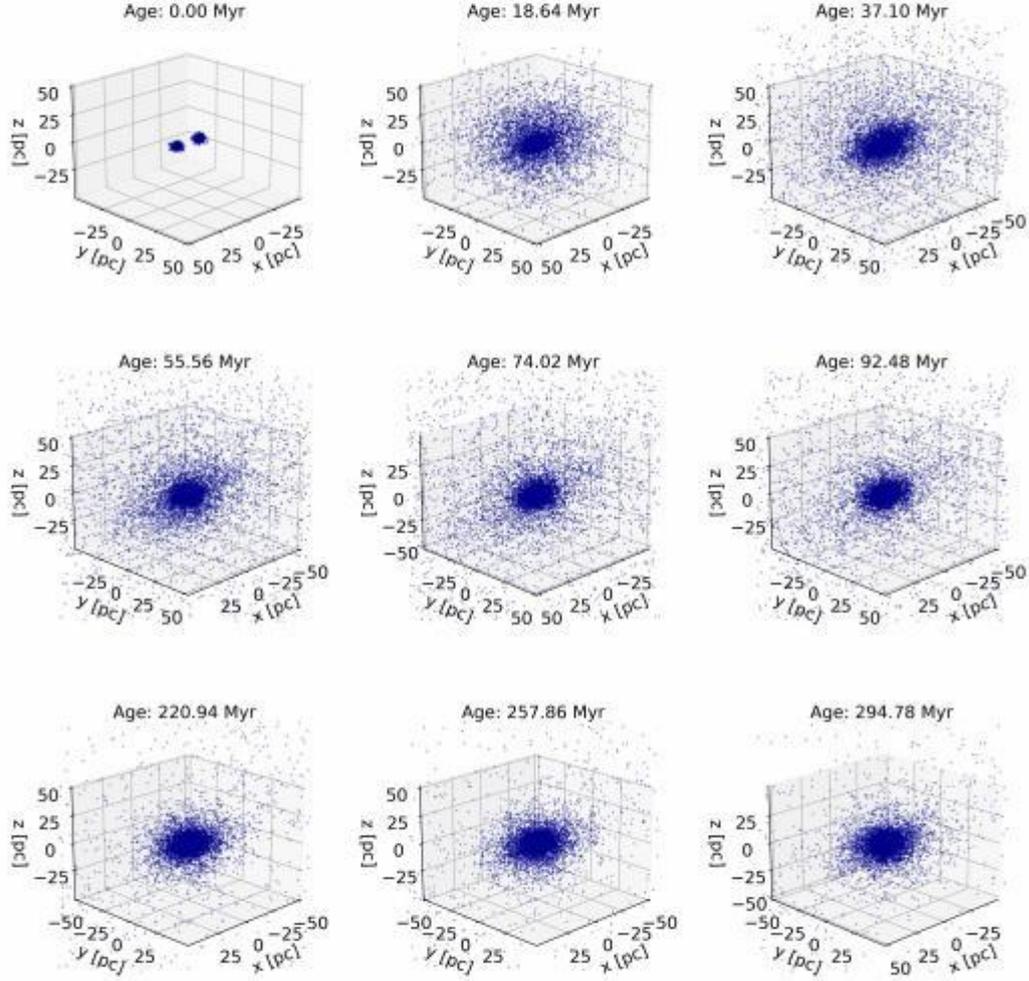

Fig. 6: Similar to Fig. 5, but for a binary cluster with initial separation of 30 pc.

Recently, Sobodar et al. (2025) also developed a multi-population N-body code based on Nbody6++GPU and studied the dynamical evolution of two components formed in one star cluster.

In addition, direct N-body simulation supplies a chance to observe the formation and evolution of stellar population features in star clusters, because the stellar evolution of both single and binary stars can be included, besides stellar motion. Some observational features such as CMDs, spectral energy distributions (SEDs) of stellar populations can be calculated from the simulation results and then compared to the observations of star clusters, using the stellar population synthesis method (Li et al. 2017b; Li & Mao 2021). In fact, stellar evolution affects the dynamical evolution of star clusters, as the mass and number of stars changed in the evolution of star clusters.

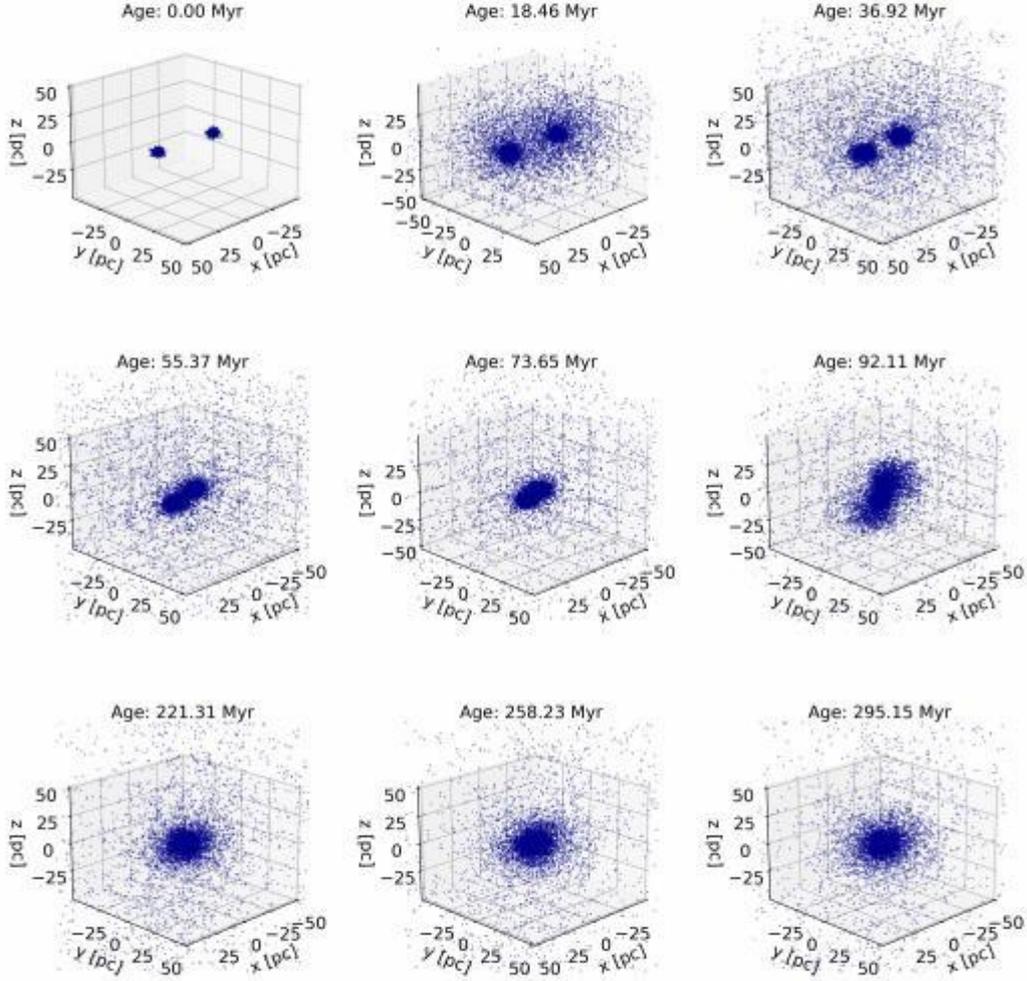

Fig. 7: Similar to Fig. 5, but for a binary cluster with initial separation of 50 pc.

**Acknowledgements** This work has been supported by the Chinese National Science Foundation (No. 12473029), Yunnan Academician Workstation of Wang Jingxiu, Dali expert workstation of Rainer Spurzem, and Guanghe Fundation (No. ghfund202407013470). Rainer Spurzem is grateful for support by the German Science Foundation (DFG), grant No. Sp 345/24-1, and acknowledges NAOC International Cooperation Office for its support in 2023, 2024, and 2025, and the support by the National Science Foundation of China (NSFC) under grant No. 12473017. We thank the Max-Planck Computing and Data Facility (MPCDF) at Garching, Germany, for computing time for the benchmarks in Fig. 8. RS thanks Thorsten Naab for frequent hospitality during visits at Max-Planck Institute for Astrophysics, Garching.

## DATA AVAILABILITY

The data of this paper can be obtained on request to Z.-M. Li.

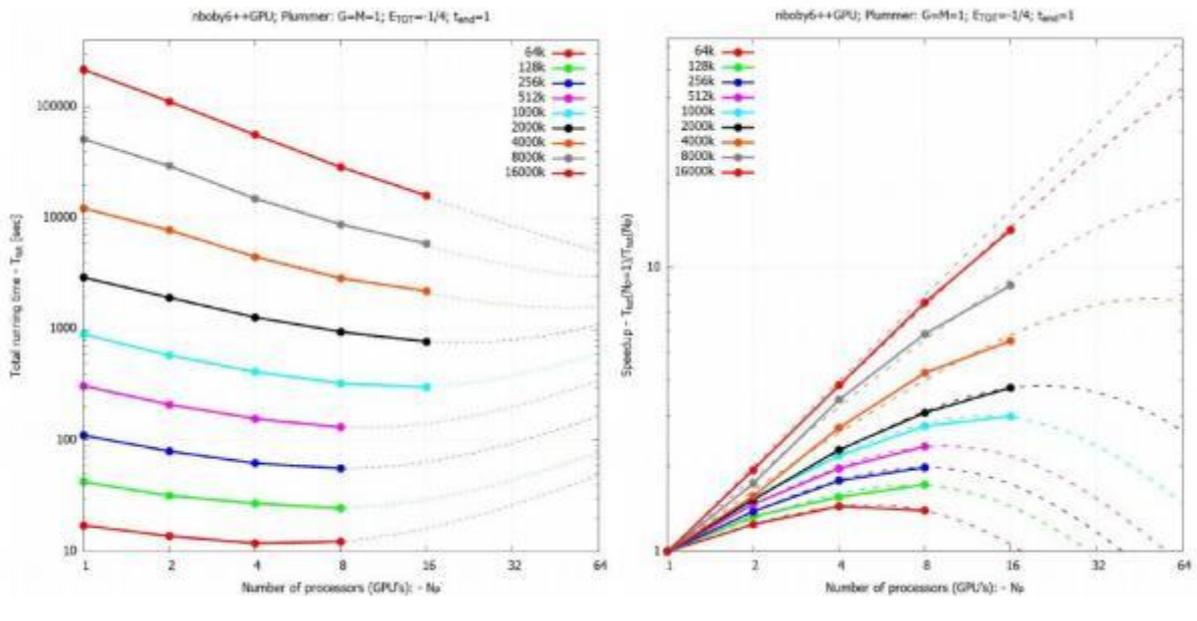

Fig. 8: Benchmark results and extrapolated scaling for NBODY 6++GPU, on the raven cluster at MPCDF, see main text. Left: Total time for one N BODY model unit in secs (order of a crossing time at half mass radius); Right: Speed-Up compared to using one GPU only. In both cases different curves for particle numbers from 64k to 16m.